\documentclass[aps,pra,aps,twocolumn,a4paper,superscriptaddress,abbrvnat,longbibliography,nofootinbib,showpacs]{revtex4-1}
\usepackage[english]{babel}
\usepackage{graphicx}
\usepackage[utf8]{inputenc}

\usepackage{amsmath}
\usepackage{amssymb}
\usepackage{xcolor}
\usepackage[colorlinks]{hyperref}

\newcommand{\bra}[1]{\langle #1|}
\newcommand{\ket}[1]{|#1\rangle}
\def\ben{\begin{eqnarray}}

\def\een{\end{eqnarray}}
\def\bei{\begin{itemize}}
	\def\eei{\end{itemize}}
\def\be{\begin{equation}}
\def\ee{\end{equation}}

\usepackage{bm}
\def\ot{\otimes}
\def\<{\langle}
\def\>{\rangle}

\def\[{\left[}
\def\]{\left]}
\usepackage{mciteplus}

\begin{document}
	\title{Decoherence of spin registers revisited}

	\author{Jan Tuziemski}\email{jan.tuziemski@pg.edu.pl}
	\affiliation{Faculty of Applied Physics and Mathematics, Technical University of Gda\'{n}sk, 80-233 Gda\'{n}sk, Poland}
	\affiliation{National Quantum Information Center of Gda\'{n}sk, 81-824 Sopot, Poland}
	\affiliation{Department of Physics, Stockholm University, Stockholm SE-106 91 Sweden}
	
		\author{Aniello Lampo}
	\affiliation{ICFO -- Institut de Ci\`encies Fot\`oniques, The Barcelona Institute of Science and Technology, 08860 Castelldefels (Barcelona), Spain}
	
	\author{Maciej Lewenstein}%\email{maciej.lewenstein@icfo.es}
	\affiliation{ICFO -- Institut de Ci\`encies Fot\`oniques, The Barcelona Institute of Science and Technology, 08860 Castelldefels (Barcelona), Spain}
	\affiliation{ICREA, Psg. Lluis Companys 23, E-08010 Barcelona, Spain}
	\author{Jaros\l aw K. Korbicz}

    %  \affiliation{Center for Theoretical Physics, Polish Academy of Sciences, 02-668, Warszawa, Poland}
	%\affiliation{Faculty of Applied Physics and Mathematics, Technical University of Gda\'{n}sk, 80-233 Gda\'{n}sk, Poland}
	%\affiliation{National Quantum Information Center of Gda\'{n}sk, 81-824 Sopot, Poland}
	\affiliation{Center for Theoretical Physics, Polish Academy of Sciences, Aleja Lotników
		32/46, 02-668 Warsaw, Poland}
\begin{abstract}
We revisit decoherence process of a multi-qubit register interacting with a thermal bosonic bath. We generalize the previous studies by considering not only the register's behavior but also of a part of its environment. In particular, we are interested
in information flow from the register to the environment, which we describe using recently introduced multipartite quantum state structures called Spectrum Broadcast Structures. Working in two specific cases of: i) two-qubit register and  ii) collective decoherence,
we identify the regimes  where the environment acquires almost complete information about the register state. We also study in more detail
the interesting causal aspects, related to the finite propagation time of the field disturbances between the qubits. Finally, we describe quantum state structures which appear due to the presence of  protected spaces.

\end{abstract}
	
	\pacs{03.67.Hk, 03.67.Mn, 03.65.Ta, 02.50.Ga}
	\date{\today}
	\maketitle
	\section{Introduction}

Decoherence of qubit registers due to an interaction with a thermal bath is a seemingly well studied process \cite{PSE1996,RQJ2002}  with all the relevant time-scales and protected spaces identified. The importance of such studies lies in quantum technological applications, and in the general understanding of the foundations of quantum physics. However, only the register's dynamics was studied with the environment treated merely as the source of noise. On the other hand, it emerges from a recent studies  under the names of quantum Darwinism \cite{ZurekNature} and Spectrum Broadcast Structures (SBS) \cite{Korbicz2014, myPRA} that, during the decoherence, the environment can gain valuable information about the system and play a role of a communication channel. This role is of a great importance for the understanding of the  quantum-to-classical transition, touching such deep questions as that of objectivity (see e.g. \cite{ZurekNature,myPRA}). Recently, environment as  a communication channel has been studied for a single qubit interacting with a thermal bath (the spin-boson model) \cite{my}. In particular, the regimes of SBS formation were identified and the relation to non-Markovianity analyzed.  In this work, we complement those studies with a similar analysis of a spin register in a thermal bosonic bath (see \cite{Mironowicz2017} for the studies of a spin environment).  It can also be regarded as a generalization of the previous works \cite{PSE1996,RQJ2002} to include (a part of) the environment. 

Generalization from a single qubit to a multi-qubit register,  in which the register
	qubits do not directly interact with each other, brings some remarkable qualitative changes. First of all, as it is well known there appear so called Decoherence Free Subspaces (DFS), which are protected sub-spaces of the register, immune to decoherence. When the environment is included, there appears a complementary notion of so called Orthogonalization Free Subspaces (OFS), for which the environment information gain is zero \cite{Mironowicz2017}. We show here an example of a simultaneous DFS and OFS, which has some non-trivial consequences for the joint system-environment state. Second, spacial separation of qubits introduces new effects, corresponding to a non-zero time-of-flight of the bosonic field disturbances between the physical locations of qubits. Known for a long time for the register qubits \cite{Milonni, RzazewskiZakowicz,   RQJ2002}, here they are for the first time studied for the environment and from the quantum information perspective. In particular, studying a two-qubit register we show that a decohering/recohering impulse felt by the register is accompanied by a similar information gain/loss impulse in fragments of the environment.   
	
The physics discussed in this paper is very much related to the physics of Dicke's superradiance \cite{Dicke54}. In particular, collective effects occurring in multi-qubit registers correspond directly to superradiant effect and radiation trapping effects. More concretely, when two emitters are close one to another and their dipoles oscillate in phase, the constructive interference leads to superradiance, i.e. the effect that the radiation rate is twice as big as the rate for a single emitter. Conversely, if the dipoles oscillate in anti-phase, the destructive interference takes place, the radiation rate goes to zero, and the radiation is trapped. In the decoherence language, this corresponds to a formation of a  Decoherence Free Subspace. The effects of constructive and destructive interference are still present, when the emitters are separated. The photon emitted by one emitter affect the second one in an constructive/destructive manner, depending if their dipole moments are in phase/out of phase, respectively.

The main tool used here to study information flow are so called Spectrum Broadcast Structures, introduced in \cite{Korbicz2014, myPRA}. Assuming that some fraction of the environment, called $fE$ is left for observation, SBS are the following multipartite quantum state structures between the central system $S$ and $fE$:
\begin{eqnarray}\label{SBS}
&&\rho_{S:fE}=\sum_ip_i\ket{x_i}\bra{x_i}\otimes\rho^{E_1}_i...\otimes\rho^{E_{fN}}_i,\\
&&\rho^{E_k}_i\perp \rho^{E_k}_{i'} \textrm{ for every } i'\ne i \textrm{ and } k={1,\dots,fN}.
\end{eqnarray}
Here $\{\ket{x_i}\}$ is the so-called pointer basis of the register to which it decoheres, $p_i$ are initial pointer probabilities,
and $\rho^{E_k}_i$ are some states of the observed parts of the environment with mutually orthogonal supports for different pointer index $i$.
The state (\ref{SBS}) describes redundantly stored information about the system, the index $i$, in the environment. Because of that, it corresponds to a certain, operational form of
objectivity of the central system's state \cite{myPRA}. It has been shown to appear in a variety of models such as the illuminated sphere model   \cite{Korbicz2014}, the spin-boson model \cite{my}, the quantum Brownian motion model \cite{Tuziemski2015,*Tuziemski2015b,*Tuziemski2016}, a simplified quantum electrodynamics model \cite{myQED}, in a recently proposed mechanism of gravitational decoherence \cite{myGrawitacja}
as well as in generic von Neumann measurements \cite{wyPomiary}. The structure (\ref{SBS}) is an idealized structure and in realistic situations one can expect only some form of an approach to it. This approach has been characterized mathematically
in \cite{MironowiczPRL} (see also \cite{Korbicz2014}) in terms of two quantities: The usual decoherence factors and state fidelities \cite{FG1999} between the environmental states $\rho^{E_k}_i$. These are the central quantities of our analysis.

The work is organized as follows. In Section \ref{s1} we recall the register model and its dynamics. In Section \ref{s2} we analyze the structure of the partially reduced state $\rho_{S:fE}$ in the model and
derive general expressions for the decoherence and fidelity factors, including full analytical solutions for both assuming the whole frequency spectrum of the environment is taken into the account.
Section \ref{s4} is dedicated to the simplest, non-trivial case, a two-qubit register.   In Section \ref{s5} we consider another simplified situation - so called collective decoherence, corresponding to very short transit times of the bosonic field perturbation compared
to the other timescales of the model. The conditions for protected subspaces are derived and the consequences for the form of the partially traced state analyzed.
Concluding remarks are presented in Section \ref{s6}. In Appendix \ref{sec:appendixa} we present an analytical derivation of the decoherence factor and state fidelities for uncut environment. We discuss the relation to the Dicke model and the related papers on superradiance and radiation trapping in the Appendix \ref{sec:appendixb}.

	\section{Model and its dynamics}\label{s1}
	We study  the
	model of a $L-$qubit register interacting with a bosonic environment. The system is described by the following Hamiltonian \cite{PSE1996,RQJ2002}:
	\ben
	H=H_S+H_E+H_{int},
	\een
	where the free dynamics of the  register and the environment is given by:
	\ben
	&&H_S=\sum_{n=1}^L J^{(n)}_z, \; \; H_E=\sum_{\boldsymbol{k}}\omega_{\boldsymbol{k}} a^{\dagger}_{\boldsymbol{k}} a,
	\een
	with  $J_z^\equiv \frac{1}{2}\sigma_z$, being the Pauli $\sigma_z$ operator acting on the $n$-th register qubit. The interaction between the  qubits and the environment modes is of a form
	\ben
	 H_{int}=\sum_{n=1}^L J^{(n)}_z \ot \sum_{\boldsymbol{k}}\left(g^n_{\boldsymbol{k}} a^\dagger_{\boldsymbol{k}} + g^{n*}_{\boldsymbol{k}} a_{\boldsymbol{k}}\right).
	\een
This kind of interaction appears naturally when one considers an ensemble of two level atoms coupled to the electromagnetic (EM) field. Usually in such systems, the free Hamiltonian of atoms is described by the sum of the $\sigma_z$ matrices, describing projections on the ground and excited states, multiplied by the corresponding energies. The dipolar  coupling with the EM field is then described by $\sigma_x$ or $\sigma_y$ matrices. In the special situations when the ground and excited states are degenerated, the free Hamiltonian vanishes, and we have to "rotate" the interaction term, so that it contains the diagonal $\sigma_z$ matrices.

	Since $[J^{(n)}_z,H]=0$, there is no energy transfer between the register and the environment - dissipation is thus not taken into account. This means that our approach is valid as long as the dissipation timescale is much larger than timescales of processes that we are interested in, what is usually the case. To derive the evolution operator it is convenient to work in the interaction picture, with interaction Hamiltonian given by:
	\ben
	H^{I}_{int}(t)= \sum_{n}  J^{(n)}_z \ot \sum_{\boldsymbol{k}} \left(g^n_{\boldsymbol{k}} a^\dagger_{\boldsymbol{k}} e^{i \omega_{\boldsymbol{k}} t} + g^{n*}_{\boldsymbol{k}} a_k e^{-i \omega_{\boldsymbol{k}} t}\right).
	\een
From the above expression we can easily derive the unitary evolution of the whole system $\hat U^I_{S:E}(t)$, using e.g. the Magnus expansion. To present the results we introduce the following notation (cf. \cite{my}):
 i) the register state is determined by a bit string vector $\boldsymbol{\epsilon} \equiv (\epsilon_1, \ldots,\epsilon_L)$, where $\epsilon_n \equiv \pm \frac{1}{2}$;
 %i) the register state is determined by a bit string vector $\boldsymbol{\epsilon} \equiv (\epsilon_1, \ldots,\epsilon_L)$, where $\epsilon_n \equiv {0,1}$ where 0 corresponds to eigenvalue $-\frac{1}{2}$ and 1 to $\frac{1}{2}$ respectively;
  ii) for $k$-th field mode the coupling constants are also arranged into a vector $\boldsymbol{g}_{\boldsymbol{k}} \equiv (g^1_{\boldsymbol{k}}, \ldots,g^L_{\boldsymbol{k}})$.  We arrive at:
\ben
&&\hat U^I_{S:E}(t)= \sum_{\boldsymbol{\epsilon}}| \boldsymbol{\epsilon} \> \< \boldsymbol{\epsilon} | \ot \bigotimes_{\boldsymbol{k}} \hat U^I_{\boldsymbol{k}} (t;\boldsymbol{\epsilon} ) \\&&
\hat U^I_{\boldsymbol{k}} (t;\boldsymbol{\epsilon} )\equiv \hat D\left( \alpha_{\boldsymbol{k}}(t)  \boldsymbol{\epsilon} \cdot \boldsymbol{g}_{\boldsymbol{k}}  \right)  e^{i \left| \boldsymbol{\epsilon} \cdot \boldsymbol{g}_{\boldsymbol{k}}  \right|^2 \xi_{\boldsymbol{k}} (t) }, \label{eq:controlunitary}\\
&& \alpha_{k}(t) \equiv  \frac{1-e^{i \omega_{\boldsymbol{k}}  t}}{\omega_{\boldsymbol{k}} }, \\
&& \xi_{\boldsymbol{k}} (t) \equiv  \frac{\omega_{\boldsymbol{k}} t-\sin \left(\omega_{\boldsymbol{k}}  t\right)}{\omega_{\boldsymbol{k}} ^2}.
\een
Above, $\hat D\left( \beta_{\boldsymbol{k}} \right) \equiv e^{ \beta_{\boldsymbol{k}} a^\dagger_{\boldsymbol{k}} - \beta_{\boldsymbol{k}}^* a_{\boldsymbol{k}}  }$ is the multimode displacement operator. In accord with the quantum Darwinism/Spectrum Broadcasting scenario we are interested in a situation where some of environmental degrees of freedom are left for observation, while other pass unobserved and hence can be traced out.

\section{The structure of the partially reduced state - general considerations}\label{s2}
Our main object of study is thus what we call a partially reduced state:
\be
\rho_{S:fE}(t)=\textrm{tr}_{(1-f)E}\rho_{S:E}(t),
\ee
where we denote symbolically by $fE$ the observed fraction of the environment, consisting of $fN$ modes, $0<f<1$, $(1-f)E$ represents the unobserved fraction of $(1-f)N$ modes, and $\rho_{S:E}(t)=\hat U_{S:E}(t)\rho_{S:E}(0)\hat U_{S:E}(t)$ is the evolved state of the full $S:E$ system. At this moment we leave the fractions $fE$ and $(1-f)E$ unspecified and will study how the partially traced state depends on them in what follows.

Assuming the usual fully product initial state $\rho_{S:E}(0)=\rho_{0S}\otimes \bigotimes_{\boldsymbol k}\rho_{0\boldsymbol k}$,
a quick calculation gives:
\ben
\label{eq:prs}
&&\rho^I_{S:fE}(t)=\sum_{\boldsymbol{\epsilon}} c_{\boldsymbol{\epsilon}\boldsymbol{\epsilon}} | \boldsymbol{\epsilon} \> \< \boldsymbol{\epsilon} | \ot \bigotimes_{\boldsymbol{k}}^{fN} \rho^I_{\boldsymbol{k}}(t; \boldsymbol{\epsilon}) + \\ &&\sum_{\boldsymbol{\epsilon}}  \sum_{\boldsymbol{\epsilon}' \neq \boldsymbol{\epsilon}} \gamma_{ \boldsymbol{\epsilon} \boldsymbol{\epsilon}'}(t) c_{\boldsymbol{\epsilon}\boldsymbol{\epsilon'}} | \boldsymbol{\epsilon} \> \< \boldsymbol{\epsilon}' | \ot  \bigotimes_{\boldsymbol{k}}^{fN} U^I_{f}(t;\boldsymbol{\epsilon} ) \rho_{0\boldsymbol{k}} \hat U^I_{f}(t;\boldsymbol{\epsilon}' )^\dagger, \nonumber
\een
where $c_{\boldsymbol{\epsilon}\boldsymbol{\epsilon'}}  \equiv \< \boldsymbol{\epsilon} | \rho_{0S} | \boldsymbol{\epsilon}' \>$,
\ben
&&\rho^I_{\boldsymbol{k}}(t;\boldsymbol{\epsilon} ) \equiv \hat U^I_{\boldsymbol{k}}(t;\boldsymbol{\epsilon} ) \rho_{0 \boldsymbol{k}} \hat U^{I}_{\boldsymbol{k}}(t;\boldsymbol{\epsilon} )^{\dagger}, \label{signal_states}\\
&&\gamma_{ \boldsymbol{\epsilon}\boldsymbol{\epsilon}'}(t)\equiv \prod_{\boldsymbol k\in (1-f)E}tr\left[\hat U^I_{\boldsymbol{k}} (t;\boldsymbol{\epsilon} )\rho_{0\boldsymbol k}\hat U^I_{\boldsymbol{k}} (t;\boldsymbol{\epsilon}' )\right]
\een
the last being the decoherence factor responsible for suppression of the register's off-diagonal terms in the $|0,1\rangle^{\otimes L}$ basis, serving  here as the register's pointer basis. Assuming that the environment is initially in a thermal state, the decoherence factor can be compactly written in a matrix form as:
\begin{eqnarray}
\label{eq:decmat}
&&-\log \gamma_{ \boldsymbol{\epsilon}\boldsymbol{\epsilon}'}(t) =\Delta \boldsymbol{\epsilon}^T \boldsymbol{\Gamma}(t) \Delta \boldsymbol{\epsilon} +\\
&&i \left[\boldsymbol{\epsilon}^T \boldsymbol{\Gamma}^+(t) \boldsymbol{\epsilon} - \boldsymbol{\epsilon}'^T \boldsymbol{\Gamma}^+(t) \boldsymbol{\epsilon}' - 2 \boldsymbol{\epsilon}^T \boldsymbol{\Gamma}^-(t) \boldsymbol{\epsilon}'\right], \nonumber
\end{eqnarray}
where $\Delta \boldsymbol{\epsilon} \equiv \boldsymbol{\epsilon} - \boldsymbol{\epsilon}'=(\epsilon_1-\epsilon'_1,\dots,\epsilon_L-\epsilon'_L)$ is vector of the differences.
We note that unlike in the single qubit case \cite{SchlosshauerBook, BreuerBook}, here the decoherence factor has both real and imaginary parts \cite{PSE1996,RQJ2002}.
Clearly, the vanishing of  the real-phase part:
\be\label{realG}
-\log \Gamma_{ \boldsymbol{\epsilon}\boldsymbol{\epsilon}'}(t) \equiv \Delta \boldsymbol{\epsilon}^T \boldsymbol{\Gamma}(t) \Delta \boldsymbol{\epsilon}
\ee
implies a decay of the off-diagonal elements and in what follows we will study this part.
To further specify the above matrices, we assume a wavelike position-dependent form of the coupling,
reflecting the wavelike character of the
bosonic modes. This can be thought as e.g. assuming that the  interactions of the register with the EM field can be well described using  the dipole approximation, where the coupling depends only on the positions of the register qubits, but not on their internal electronic degrees of freedom:
\ben
\boldsymbol{g}_{\boldsymbol{k}}  = g_k  \left(e^{-i\boldsymbol{k} \boldsymbol{r}_1}, \ldots,e^{-i\boldsymbol{k} \boldsymbol{r}_L} \right).
\een
Then the elements of matrices entering Eq. (\ref{eq:decmat}) are given by \cite{PSE1996,RQJ2002}:
\ben
&&\boldsymbol{\Gamma}_{nm}(t) \equiv \frac{1}{2} \sum_{\boldsymbol{k}  \in (1-f)E} |g_k  \alpha_{\boldsymbol{k}} (t) |^2  \coth \left(\frac{\omega_{\boldsymbol{k}} }{2k_BT}\right) \cos \left(\boldsymbol{k} \Delta \boldsymbol{r}_{nm}\right) \nonumber \\ \label{G0}\\ %\frac{1-\cos \omega_{\boldsymbol{k}}  t}{\omega_{\boldsymbol{k}} ^2} \\
&&\boldsymbol{\Gamma}^+_{nm}(t) \equiv\sum_{\boldsymbol{k}  \in (1-f)E} |g_k|^2 \xi_{\boldsymbol{k}} (t)  \cos \left(\boldsymbol{k} \Delta\boldsymbol{r}_{nm}\right),  \label{G0+}\\ %\frac{\omega_{\boldsymbol{k}} t - \sin \omega_{\boldsymbol{k}}  t}{\omega_{\boldsymbol{k}} ^2}
&&\boldsymbol{\Gamma}^-_{nm}(t) \equiv\frac{1}{2}\sum_{\boldsymbol{k}  \in (1-f)E} |g_k \alpha_{\boldsymbol{k}} (t)|^2 \sin \left(\boldsymbol{k} \Delta\boldsymbol{r}_{nm}\right) ,\label{G0-}
\een
with $\Delta \boldsymbol{r}_{nm}= \boldsymbol{r}_n - \boldsymbol{r}_m$ being the physical distance between the register qubits.

The novelty of our approach, compared to the standard treatments \cite{PSE1996,RQJ2002} is that we are interested not only in the state of the register, but also in the part of its environment.  Especially we will be  interested in what, if any, system-related information those observed parts of the environment obtain during the evolution.
As a measure of the information content we will choose the state fidelity of the  states (\ref{signal_states}) for different  $\boldsymbol \epsilon$, $\boldsymbol \epsilon'$ \cite{Korbicz2014, my}:
\be
B^{(\boldsymbol k)}_{\boldsymbol{\epsilon}\boldsymbol{\epsilon'}}(t) \equiv tr \sqrt{\sqrt{\rho_{\boldsymbol k}(t;\boldsymbol{\epsilon})}\rho_{\boldsymbol k}(t;\boldsymbol{\epsilon'}) \sqrt{\rho_{\boldsymbol k}(t;\boldsymbol{\epsilon})}}.\label{microB}
\ee
Just like in the previous studies \cite{Korbicz2014, my}, we will be interested in some sort of a thermodynamic limit with large $N$ and the information content of macroscopic groups of modes rather than of single modes which may contain vanishingly small information about the register \cite{Korbicz2014}.  We will thus divide the observed modes into bigger groups - called macrofractions $\cal M$ \cite{Korbicz2014}. %From Eq. (\ref{eq:prs}) we see that the environment constitutes of a displaced thermal states.
We define the observed fraction of the environment to be $fE \equiv \cup^{fM}_{{\cal M}=1} mac_{\cal M}$ with a state of a macrofraction simply defined as:
\ben
\rho_{\cal M}(t;\boldsymbol{\epsilon}) = \bigotimes_{\boldsymbol{k} \in \cal M} \rho_{\boldsymbol{k}}(t, \boldsymbol{\epsilon}).
\een
The quantity we will be interested in is thus a macrofraction overlap rather than the microscopic one (\ref{microB}):
\begin{eqnarray}
&&B^{{\cal M}}_{\boldsymbol{\epsilon}\boldsymbol{\epsilon'}}(t) \equiv tr \sqrt{\sqrt{\rho_{\cal M}(t;\boldsymbol{\epsilon})}\rho_{\cal M}(t;\boldsymbol{\epsilon'}) \sqrt{\rho_{\cal M}(t;\boldsymbol{\epsilon})}}\\
&&=\prod_{\boldsymbol k \in \mathcal M}tr \sqrt{\sqrt{\rho_{\boldsymbol k}(t;\boldsymbol{\epsilon})}\rho_{\boldsymbol k}(t;\boldsymbol{\epsilon'}) \sqrt{\rho_{\boldsymbol k}(t;\boldsymbol{\epsilon})}}=
\prod_{\boldsymbol k \in \mathcal M}B^{(\boldsymbol k)}_{\boldsymbol{\epsilon}\boldsymbol{\epsilon'}}(t)\nonumber
\end{eqnarray}
In the considered model the above overlap can be calculated and reads:
\ben\label{fidelity}
- \log B^{{\cal M}}_{\boldsymbol{\epsilon} \boldsymbol{\epsilon}'}(t) =\Delta \boldsymbol{\epsilon}^T \boldsymbol B^{{\cal M}}(t) \Delta \boldsymbol{\epsilon},
\een
where we define a $L\times L$ matrix:
\ben
\boldsymbol B^{{\cal M}}_{nm}(t)\equiv \frac{1}{2}\sum_{\boldsymbol{k}  \in {\cal M}} |g_k \alpha_{\boldsymbol{k}} (t)|^2 \tanh \left(\frac{\omega_{\boldsymbol{k}} }{2k_BT}\right) \cos \left(\boldsymbol{k}  \Delta \boldsymbol{r}_{nm}\right). \nonumber \label{fid}\\
\een

Functions $\Gamma_{\boldsymbol{\epsilon} \boldsymbol{\epsilon}'}$, $B^{{\cal M}}_{\boldsymbol{\epsilon} \boldsymbol{\epsilon}'}(t)$ are the basic objects of our study. If they simultaneously vanish, the partially traced state approaches
the SBS form \cite{MironowiczPRL}.
%(\ref{eq:decmat}, \ref{G}, \ref{G+}, \ref{G-}) and (\ref{fidelity},\ref{fid})
We stress that unlike in the case of a single qubit  \cite{my}, now they are  given by matrix expressions, which as we will see will lead to qualitatively different behavior.
Instead of working with relative distances and wave vectors, let us introduce transit times $\tau_{nm}$, defined as the  times that a signal needs to travel between
$n^{th}$ and $m^{th}$ qubit \cite{RQJ2002}:
\ben
\label{eq:transitt}
\boldsymbol{k} \Delta \boldsymbol{r}_{nm} \equiv \omega \tau_{nm}.\label{transit}
\een
For example, in the case of a solid state implementation of the register, the bosonic bath can be usually associated with the phonon field, so the transit time will determine speed of information propagation via phonons \cite{RQJ2002}. Then, assuming that the fractions of the environment we are working with are large, we pass to the usual continuum limit and introduce spectral density $J(\omega)$. In these terms the elements of decoherence and fidelity matrices take the form:
\ben
\label{eq:moddec}
&&\boldsymbol{\Gamma}_{nm}(t) = \label{G}\\ \nonumber &&  \int_{(1-f)E} d \omega J(\omega)  \frac{1-\cos(\omega t)}{\omega^2} \coth \left(\frac{\omega}{2k_BT}\right) \cos \left(\omega \tau_{nm}\right), \nonumber \\ %\frac{1-\cos \omega_{\boldsymbol{k}}  t}{\omega_{\boldsymbol{k}} ^2} \\
\label{eq:dec+}
&&\boldsymbol{\Gamma}^+_{nm}(t)= \label{G+}\int_{(1-f)E} d \omega J(\omega) \frac{\omega t - \sin (\omega t)}{\omega^2}  \cos \left(\omega \tau_{nm}\right),  \\ %\frac{\omega_{\boldsymbol{k}} t - \sin \omega_{\boldsymbol{k}}  t}{\omega_{\boldsymbol{k}} ^2}
\label{eq:dec-}
&&\boldsymbol{\Gamma}^-_{nm}(t) = \label{G-}\frac{1}{2}\int_{(1-f)E} d \omega J(\omega) \frac{1-\cos(\omega t)}{\omega^2} \sin \left(\omega \tau_{nm}\right) ,
\\
&& \boldsymbol B^{{\cal M}}_{nm}(t) = \label{B}\\ \nonumber &&  \int_{\cal M} d \omega J(\omega)  \frac{1-\cos(\omega t)}{\omega^2} \tanh \left(\frac{\omega}{2k_BT}\right) \cos \left(\omega \tau_{nm}\right).
\een
In the above expressions $(1-f)E$ and ${\cal M}$  denote symbolically the sets of unobserved and  observed  frequencies respectively. We note that  in each of the above matrices all the diagonal entries are equal as the transit times $\tau_{mn}$ drop out of the expressions. Moreover, matrices $\boldsymbol{\Gamma}(t)$, $\boldsymbol{\Gamma}^+(t)$, and $\boldsymbol B^{{\cal M}}(t)$ are real symmetric, while $\boldsymbol{\Gamma}^-(t)$ is real skew-symmetric. As the spectral density we adopt the usual for spin-boson models expression:
\ben
\label{sd}
J(\omega) = \frac{\omega^s}{\Lambda^{s-1}}\exp \left[-\omega/\Lambda\right],
\een
where $\Lambda$ is the cut-off frequency and $s$ the Ohmicity parameter.

There are several ways to divide environmental degrees of freedom into observed and unobserved parts \cite{my}. %and we will consider them in further parts of this Section.
Firstly, one can assume that, due to their large size, both unobserved and observed parts of the environment are described by the full spectral density. We will refer to this case as \textit{uncut} spectral density. In this case the above integrals are solvable analytically. Due to their length, the formulas are presented in Appendix \ref{sec:appendixa}. Here we will study  them numerically in further parts of the manuscript  for a two-qubit register.

The second possibility is that the observed and unobserved parts of environment are formed by a given parts of the spectrum. This situation can be pictured as an observation of the environment via a narrow band detector (rather then a wide-band as above). We will assume the observed frequencies are formed by a spectral interval $[ \alpha, \beta]$ and study dependence of the decoherence and the state fidelity on the position of the interval. This case will be referred to as \textit{cut} spectral density. Based on the studies from \cite{my}, one can assume a sharp spectral cut as there is no qualitative difference between the sharp and soft cuts.

\subsection{Decoherence and orthogonalization free subspaces}
\label{subsec:subspaces}

Depending on the values of the decoherence factor Eq. (\ref{eq:decmat})  and state fidelity Eq. (\ref{fidelity}), the structure of the partially reduced state Eq. (\ref{eq:prs}) may be in a good approximation of a SBS. However, it is also possible that, for some states of the register, one of the discussed processes will not take place. In such a case,
as in \cite{Mironowicz2017}, we will say that a subspace $S \subseteq \left\{ \pm \frac{1}{2} \right\}^L$ exhibits a strong Decoherence Free Subspace (DFS) property if:
\ben
\label{eq:sdfs}
\forall_{t \in \mathbb{R}_+}\forall_{\boldsymbol{\epsilon},\boldsymbol{\epsilon'} \in S} \gamma_{\boldsymbol{\epsilon}\boldsymbol{\epsilon'}} (t)=
1,
\een
and a weak DFS if
\ben
\forall_{t \in \mathbb{R}_+}\forall_{\boldsymbol{\epsilon},\boldsymbol{\epsilon'} \in S} \left| \gamma_{\boldsymbol{\epsilon}\boldsymbol{\epsilon'}} (t)\right| =
1.
\een
Similarly we define Orthogonalization Free Subspace (OFS) to occur when:
\ben
\label{eq:ofs}
\forall_{t \in \mathbb{R}_+}\forall_{\boldsymbol{\epsilon},\boldsymbol{\epsilon'} \in S} \forall_{{\cal M}} B^{{\cal M}}_{\boldsymbol{\epsilon}\boldsymbol{\epsilon'}}(t) =1.
\een
In general, due to the fact that expressions for decoherence factor and state fidelity are quite involved, it is not possible to analytically determine which states form DSF or OSF. However, when register's qubits interact collectively with the environment one may introduce conditions for $DSF$ and $OSF$. This case is discussed in detail in Section \ref{sec:colletive}.

\section{Two-qubit register}\label{s4}

A general study of the $L$-qubit register is quite complicated due to the matrix character of both decoherence factor and state fidelity. Here we study the first non-trivial register, consisting of two qubits, extending the original analysis of \cite{RQJ2002} from decoherence to SBS.
For this case, there is only one transit time $\tau$ and the real decoherence factor between the register states $| \boldsymbol{\epsilon} \> $ and $ |\boldsymbol{\epsilon'} \>$ reads (cf. (\ref{realG})):
%(cf. Eqs. (\ref{eq:moddec},\ref{eq:dec+},\ref{eq:dec-})):
\ben
- \log \Gamma_{\boldsymbol{\epsilon}\boldsymbol{\epsilon}'}(t) =&& ||\Delta \boldsymbol{\epsilon}||^2 \boldsymbol{\Gamma}_{11}(t)  +2\Delta  \epsilon_{1} \Delta \epsilon_{2} \boldsymbol{\Gamma}_{12}(t), % +  \nonumber \\ &&i \left[ \left(\boldsymbol{\epsilon}_1 \boldsymbol{\epsilon}_2 - \boldsymbol{\epsilon}_1' \boldsymbol{\epsilon}_2'\right) \left( \boldsymbol{\Gamma}^+_{01}(t) - \boldsymbol{\Gamma}^-_{01} \right) \right],
% \boldsymbol{\Gamma}^+_{00}(t) \left( |\boldsymbol{\epsilon}|^2 - |\boldsymbol{\epsilon} '|^2 \right)
\een
where $||\Delta \boldsymbol{\epsilon}||^2 = \left( \Delta \epsilon_{1}\right)^2+\left( \Delta \epsilon_{2}\right)^2$. We are interested here only in the real part of (\ref{eq:decmat}) as it is sufficient for showing damping of the off-diagonal elements. Similarly one finds that:
\ben
- \log B^{{\cal M}}_{\boldsymbol{\epsilon} \boldsymbol{\epsilon}'}(t) =&&  ||\Delta \boldsymbol{\epsilon}||^2 \boldsymbol B^{{\cal M}}_{11}(t) +2 \Delta \epsilon_{1} \Delta \epsilon_{2} \boldsymbol B^{{\cal M}}_{12}(t) .
\een
As a result, for a $2$-qubit register, there are three groups  of non-diagonal density matrix elements, presented in Table \ref{tab:states}, responsible for different types of coherence and decohering in a different manner accordingly to the value of $\Delta \epsilon_{1} \Delta \epsilon_{2}$. Similarly, the distinguishability of the environmental states depends on $\Delta \epsilon_{1} \Delta \epsilon_{2}$ too. We can distinguish the following non-trivial cases:

%\begin{widetext}

\begin{table*}[t]
	\begin{tabular}{ | c | c | c | }
		\hline
		$\Delta \epsilon_{1} \Delta \epsilon_{2}= 0$ & $\Delta \epsilon_{1} \Delta \epsilon_{2}= - 1$  &  $\Delta \epsilon_{1} \Delta \epsilon_{2}=  +1$  \\
		single qubit & singlet & GHZ\\ \hline
		\begin{tabular}{c c} $(|\frac{1}{2},\frac{1}{2}\>,|\frac{1}{2},-\frac{1}{2}\>)$, $(|\frac{1}{2},\frac{1}{2}\>,|-\frac{1}{2},\frac{1}{2}\>)$,  & \\ $(|\frac{1}{2},-\frac{1}{2}\>, |-\frac{1}{2},-\frac{1}{2}\>)$, $(|-\frac{1}{2},\frac{1}{2}\>, |-\frac{1}{2},-\frac{1}{2}\>)$& \end{tabular} & $(|\frac{1}{2},-\frac{1}{2}\>,|-\frac{1}{2},\frac{1}{2}\>)$ & $(|\frac{1}{2},\frac{1}{2}\>,|-\frac{1}{2},-\frac{1}{2}\>)$ \\ \hline
	\end{tabular}
	\caption{\label{tab:states}Pairs of states $(|\boldsymbol{\epsilon}\>,|\boldsymbol{\epsilon}'\>)$  accordingly to their $\Delta \epsilon_{1} \Delta \epsilon_{2}$ value. Diagonal elements $(|\boldsymbol{\epsilon}\>,|\boldsymbol{\epsilon}\>)$ are not taken into account.  }
\end{table*}
%\end{widetext}

\subsection{Effectively single qubit case}
This is the case when vectors $\boldsymbol{\epsilon}, \boldsymbol{\epsilon}'$ differ at most at one position e.g. $\epsilon_1 = \epsilon_1'$, $\epsilon_2 \neq \epsilon_2'$  or equivalently $\Delta \epsilon_{1} \Delta \epsilon_{2}=0$. This subspace is spanned by four states as shown in Table \ref{tab:states}). One obtains:
\ben
\label{eq:decs}
-\log \Gamma_{\boldsymbol{\epsilon}\boldsymbol{\epsilon}'}(t)=&& \boldsymbol{\Gamma}_{11}(t) %  + i \sign \left( \boldsymbol{\epsilon}_1 \left( \boldsymbol{\epsilon}_2 -  \boldsymbol{\epsilon}_2'\right) \right) \left(   \boldsymbol{\Gamma}^+_{01}(t) -\boldsymbol{\Gamma}^-_{01}  \right), \nonumber \\
\een
and
\ben
\label{eq:fids}
- \log B^{{\cal M}}_{\boldsymbol{\epsilon} \boldsymbol{\epsilon}'}(t) =&& \boldsymbol B^{{\cal M}}_{11}(t)  .
\een
In this case, apart from the phase, the register behaves effectively as a single spin interacting with bosonic bath. Therefore we will refer to it as "single qubit" case. This conclusion is not restricted just to a $2$-qubit register. To see this, consider two states of a $L$-qubit register that differ at $n$-th position, then it follows from Eq. (\ref{eq:decmat}) that
$\log \Gamma_{\boldsymbol{\epsilon}\boldsymbol{\epsilon}'}(t) =- \boldsymbol{\Gamma}_{nn}(t)=- \boldsymbol{\Gamma}_{11}(t)$, as all the diagonal elements are equal, and a similar result holds for fidelity. A detailed investigation of the SBS formation for a single central spin has been performed in \cite{my} and we refer the reader to that work. Here, we will use this case only as a reference to highlight novel features of the register model. For a fair comparison, we note that Hamiltonian of the spin-boson model is usually defined using $\sigma_z$ and here we used $\frac{1}{2}\sigma_z$, following the common quantum register convention. Consequently, the comparison of the result presented here with those of \cite{my} should take into account that in the spin-boson model Eqs. (\ref{eq:decs},\ref{eq:fids}) read $\log \Gamma_{\boldsymbol{\epsilon}\boldsymbol{\epsilon}'}(t) =-4\boldsymbol{\Gamma}_{11}(t)$ and $\log B^{{\cal M}}_{\boldsymbol{\epsilon} \boldsymbol{\epsilon}'}(t) =-4\boldsymbol B^{{\cal M}}_{11}(t)$, respectively.

\subsection{True $2$-qubit case}
The remaining non-trivial case is when the vectors  $\boldsymbol{\epsilon}, \boldsymbol{\epsilon}'$ differ at all positions, i.e.  $\epsilon_1 \neq \epsilon_1'$, $\epsilon_2 \neq \epsilon_2'$. This situation is described by pairs of states such that $\Delta \epsilon_{1} \Delta \epsilon_{2}=\pm 1$, see Table \ref{tab:states}. Then the corresponding expressions take a form:
\ben
\label{eq:decf2}
-\log\Gamma_{\boldsymbol{\epsilon}\boldsymbol{\epsilon}'}(t) =&& 2 \left[ \boldsymbol{\Gamma}_{11}(t) + \Delta \epsilon_{1} \Delta \epsilon_{2} \boldsymbol{\Gamma}_{12}(t)  \right],% \\   && + i \sign \left( \boldsymbol{\epsilon}_1  \boldsymbol{\epsilon}_2 - \boldsymbol{\epsilon}_1' \boldsymbol{\epsilon}_2' \right) \left(   \boldsymbol{\Gamma}^+_{01}(t) -\boldsymbol{\Gamma}^-_{01}  \right), \nonumber
\een
and
\ben
\label{eq:fidf2}
- \log B^{{\cal M}}_{\boldsymbol{\epsilon} \boldsymbol{\epsilon}'}(t) =&& 2\left[\boldsymbol B^{{\cal M}}_{11}(t) +\Delta \epsilon_{1} \Delta \epsilon_{2} \boldsymbol B^{{\cal M}}_{12}(t)  \right]. \nonumber \\
\een

Note that in the Dicke superradiant limit, when $\boldsymbol{k} \Delta \boldsymbol{r}_{nm} = \omega \tau_{nm} \to 0$, the RHS of (\ref{eq:decf2}) tends to 
$4\boldsymbol{\Gamma}_{11}(t)$ or zero, indicating, as expected,  superradiance or radiation trapping, respectively. In the latter case  there will be a strong DFS (cf. Eq. (\ref{eq:sdfs})). Similarly,  $\boldsymbol B^{{\cal M}}_{11}(t) =\boldsymbol B^{{\cal M}}_{12}(t)$ and $\Delta \epsilon_{1} \Delta \epsilon_{2} = -1$ leads to an OFS (cf. Eq. \ref{eq:ofs}). We will study such cases in more detail in Section \ref{sec:colletive} dedicated to collective decoherence.

\begin{figure}
	\begin{center}
		\includegraphics[width=1\columnwidth]{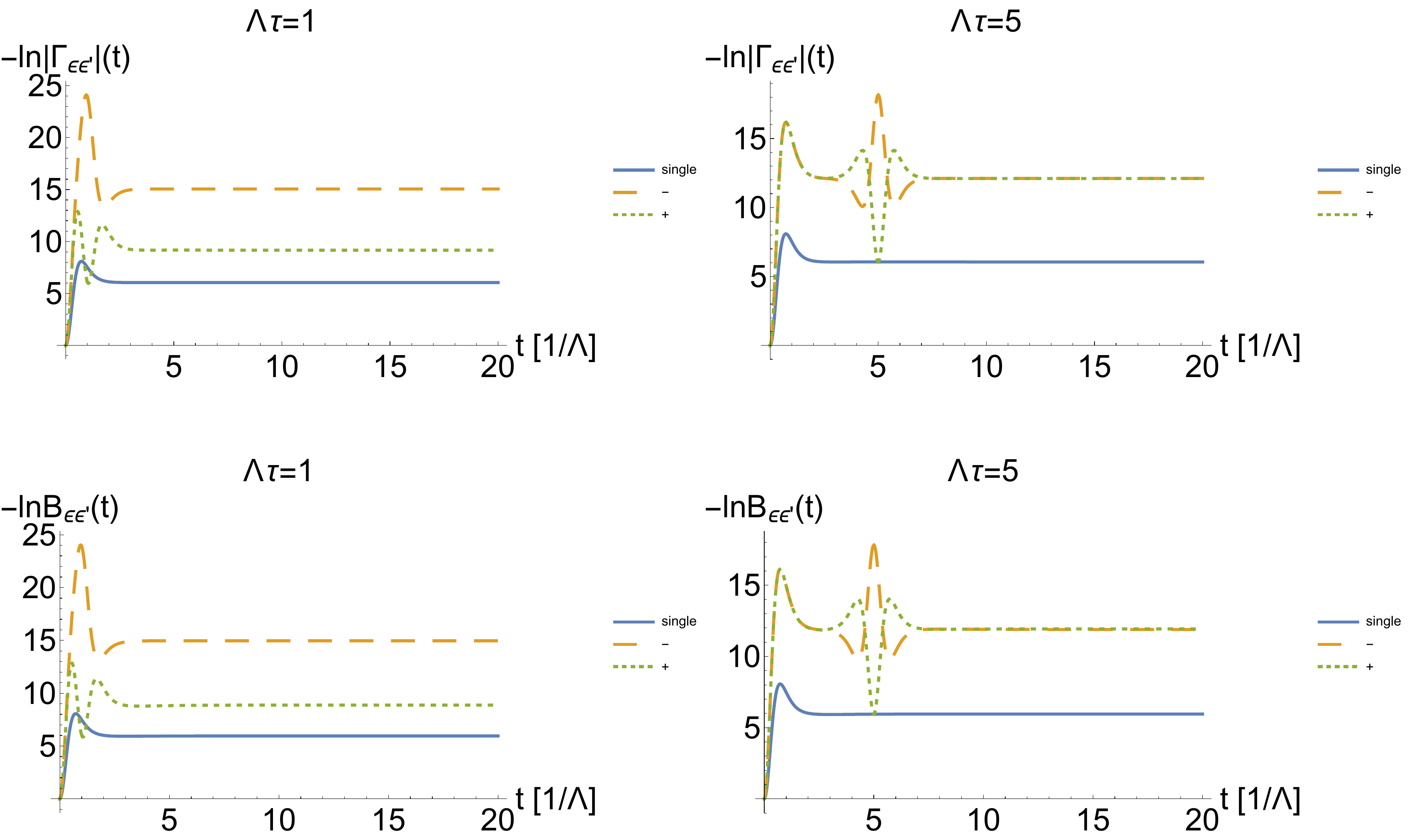}
		\caption{\label{fig:dbuncut} Uncut spectral density case, different transit times: Logarithm of decoherence factor (Eq. (\ref{eq:decf2})) -- upper trace, and fidelity (Eq. (\ref{eq:fidf2})) -- lower trace for different values  of transit time $\tau$ (Eq. (\ref{eq:transitt})): $\Lambda \tau = 1$ - upper left and lower left, $\Lambda \tau = 5$ - upper right and lower right. In each plot there are three curves corresponding to states with different values of $\Delta \epsilon_{1} \Delta \epsilon_{2}$: minus - dashed line and plus - dotted line as well as for the "single qubit" case (Eq. (\ref{eq:decs})) - solid line. In all plots $T=\Lambda/3$ and $s=5$.   }
	\end{center}
\end{figure}

\begin{figure}
	\begin{center}
		\includegraphics[width=1\columnwidth]{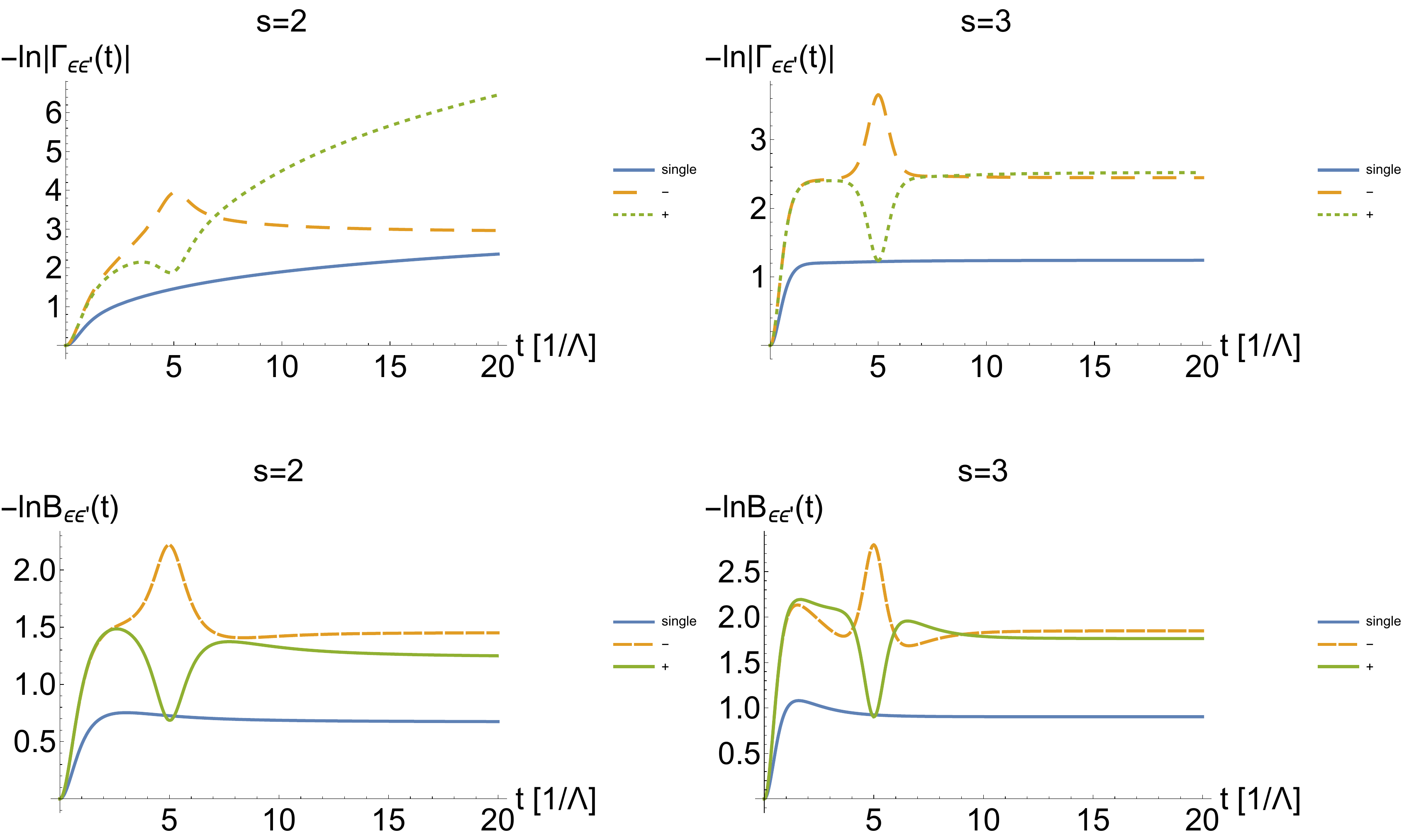}
		\caption{\label{fig:dbs} Uncut spectral density case, different ohmicity parameter: Logarithm of the decoherence factor (Eq. (\ref{eq:decf2})) -- upper trace, and fidelity (Eq. (\ref{eq:fidf2})) --lower trace for different values of Ohmicity parameter (Eq. (\ref{sd})) $s$: $s =2$ - upper left and lower left, $s = 3$ - upper right and lower right, which for a spin-boson model correspond to markovian and non-Markovian evolution \cite{Addis2014}. In each plot there are three curves corresponding to states with different values of $\Delta \epsilon_{1} \Delta \epsilon_{2}$: minus - dashed line and plus - dotted line as well as for the "single qubit" case (Eq. (\ref{eq:decs})) - solid line.  In all plots $T=\Lambda/3$ and $\Lambda \tau=5$.  }
	\end{center}
\end{figure}

\begin{figure}
	\begin{center}
		\includegraphics[width=1\columnwidth]{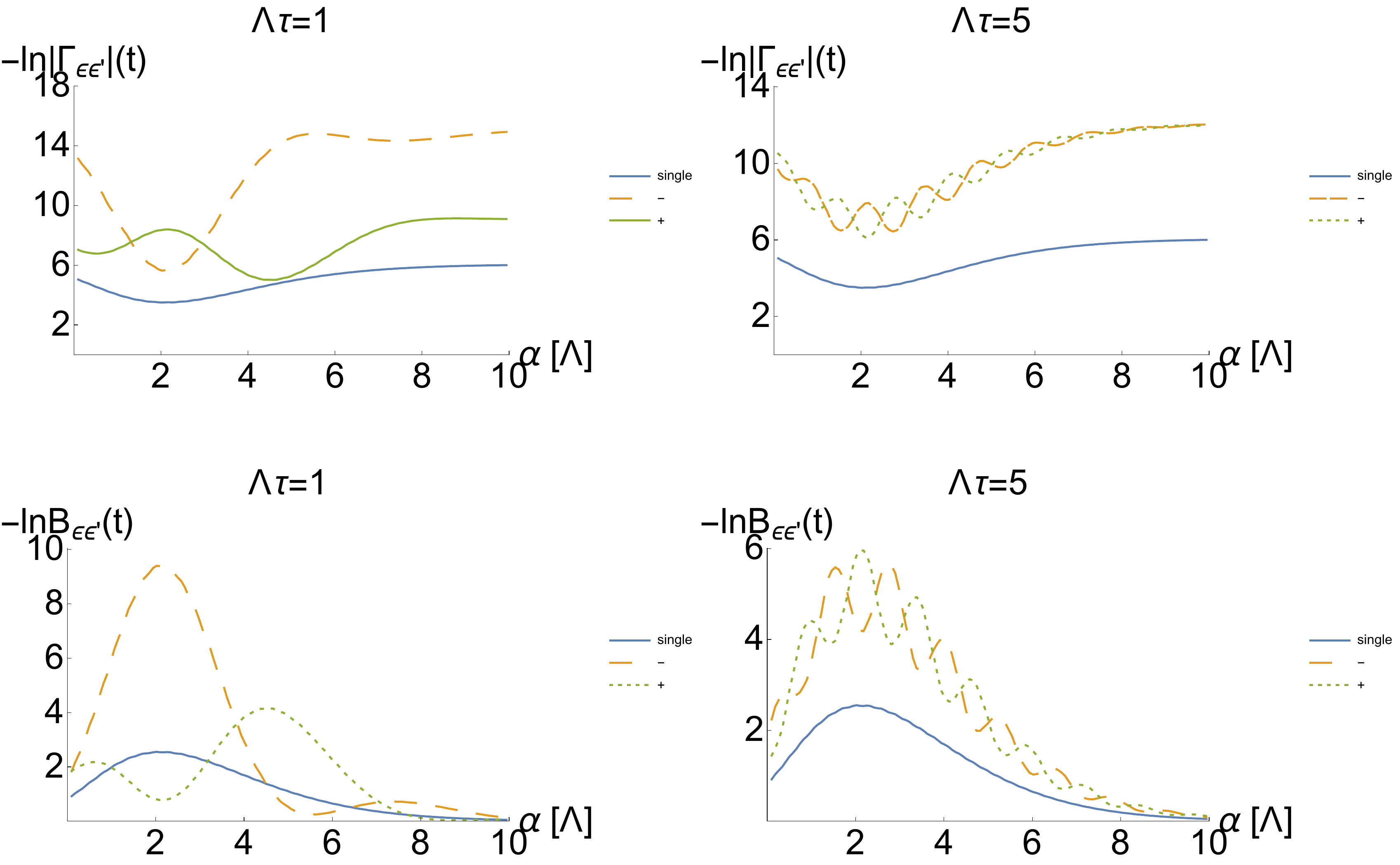}
		\caption{\label{fig:dbcut} Cut spectral density case: Time-asymptotic value of the decoherence factor (Eq. (\ref{eq:decf2})) -- upper trace, and fidelity (Eq. (\ref{eq:fidf2})) -- lower trace as a function of the cut for different values of transit time $\tau$ (Eq. (\ref{eq:transitt})): $\Lambda \tau = 1$ - upper left and lower left, $\Lambda \tau = 5$ - upper right and lower right.   In each plot there are three curves corresponding to states with different values of $\Delta \epsilon_{1} \Delta \epsilon_{2}$: minus - dashed line and plus - dotted line as well as for the "single qubit" case (Eq. (\ref{eq:decs})) - solid line. The unobserved frequencies belong to $(0, \alpha / \Lambda) \cup ((\alpha + \Delta)/\Lambda, \infty)$, whereas the observed ones to  $[\alpha / \Lambda, (\alpha + \Delta)/\Lambda]$. In all plots $T=\Lambda/3$, $s=5$ and $\Delta =2$.  }
	\end{center}
\end{figure}

We perform further studies of  (\ref{eq:decf2},\ref{eq:fidf2})  numerically. Although for the uncut case we have the analytical formulas in Appendix \ref{sec:appendixa}, it is more convenient to plot them.
In Fig. \ref{fig:dbuncut} we present results for the uncut spectral density - the decoherence factor and the fidelity for different values of the transit time $\tau$, rescaled to the cut-off $\Lambda$ and assuming a moderate temperature $0\ll T\ll s\Lambda$.
In each plot there are two curves corresponding to different values of $\Delta \epsilon_{1} \Delta \epsilon_{2}$, and  the "single qubit" case (cf. (\ref{eq:decs}, \ref{eq:fids})) for a comparison.
One immediately sees that both decoherence and the information gain by the environment are more efficient for a register than for a single qubit.
More importantly, there is also a qualitatively new behavior here: Around the transit time $t=\tau$ there appears a characteristic disturbance in both the decoherence and fidelity plots. Weather it is a dip or a peak depends on the parity of  $\Delta \epsilon_{1} \Delta \epsilon_{2}$: In the superradiant case ($\Delta \epsilon_{1} \Delta \epsilon_{2}=+1$) it corresponds to  a dip in the plotted curve, but a peak in the function $ \Gamma_{\boldsymbol{\epsilon}\boldsymbol{\epsilon}'}(t)$ (we plot $-\log \Gamma_{\boldsymbol{\epsilon}\boldsymbol{\epsilon}'}(t)$) -- as expected in this case the signal comes ``in phase''. Conversely, 
  in the radiation trapping case ($\Delta \epsilon_{1} \Delta \epsilon_{2}=-1$) it corresponds to  a peak in the curve, but a dip in the function $ \Gamma_{\boldsymbol{\epsilon}\boldsymbol{\epsilon}'}(t)$ -- as expected in this case the signal comes ``in anti- phase''.  This behavior is due to a relaxation process, where the qubits exchange a quantum of the bosonic field after the interaction has been switched on. It is basically a simpler version of the multiple retardation effects studied in the full model, with non-trivial qubit dynamics (see e.g. \cite{Milonni,RzazewskiZakowicz}). 
 There are no multiple signals here due to the trivial qubit Hamiltonian.

The disturbance is described by the vacuum part (see Appendix A), common to both decoherence and fidelity factors. More precisely, by  the second term in (\ref{Gvac}):
\be
- \frac{\cos \left[(s-1) \arctan \left(\Lambda (t-\tau)\right)\right]}{\left[1+\Lambda^2(t-\tau)^2)\right]^{\frac{s-1}{2}} }.
\ee
This term describes also the smaller disturbances, surrounding the main one in time, which appear for higher Ohmicity parameters $s$ and $\Lambda\tau$ due to the cosine periodicity (see Fig. \ref{fig:dbuncut} for $\Lambda\tau=5$). These disturbances indicate the breakdown of causality in our model for the time $\Lambda \tau\simeq 1$. As discussed for instance in \cite{Milonni,RzazewskiZakowicz}, it is due to the introduction of the cut and causality is restored for longer times, or for all quantities for which the limit $\Lambda \to \infty$ has a mathematical sense.  
%A mathematical manifestation of this process is that for $t=\tau$  the arguments of the polygamma functions entering the exact expressions for decoherence and state fidelity (Eqs. (\ref{eq:decan}) and (\ref{eq:fidan})) become real. }

The peaks in the plots for  $\Delta \epsilon_{1} \Delta \epsilon_{2}=-1$, i.e. for the pair of states $|\frac{1}{2},-\frac{1}{2}\>,|-\frac{1}{2},\frac{1}{2}\>$, imply that around $t=\tau$  both the decoherence factor and the overlap become small for this pair. This in turn implies that the partial state's projection onto the subspace spanned by $|\frac{1}{2},-\frac{1}{2}\>,|-\frac{1}{2},\frac{1}{2}\>$ approaches SBS \cite{Mironowicz2017} much better than at other times (a "blink of objectivity"):
\ben
&&\rho^I_{S:fE}(t=\tau) \approx \sum_{\boldsymbol{\epsilon}=+-,-+ } c_{\boldsymbol{\epsilon}\boldsymbol{\epsilon}}| \boldsymbol{\epsilon} \> \< \boldsymbol{\epsilon} | \otimes \bigotimes_{\cal M} \rho^I_{\cal M}(\tau; \boldsymbol{\epsilon}) + \textrm{rest},\nonumber\\
\een
and $\rho^I_{\cal M}(\tau; 01)$ and $\rho^I_{\cal M}(\tau; 10)$ have a very small overlap.
%(cf. Figs. \ref{fig:dbs}, \ref{fig:dbuncut}).  Thus,  there is a "blink of objectivity".
In contrary, the dips for  $\Delta \epsilon_{1} \Delta \epsilon_{2}=+1$, signify that both decoherence and the overlap functions temporarily rise, indicating a departure from SBS and a form of a transient recoherence. As a result, the partially traced state has a rather complicated structure around $t=\tau$, with parts well approximated by SBS and parts with restored quantum correlations. It thus in a sense simultaneously possesses classical (SBS) and quantum (coherences) properties.

In Fig.~\ref{fig:dbuncut} we can also see the influence of the transit time.  For low transit times compared to the cut-off time-scale, both decoherence and fidelity curves split  with the parity of $\Delta \epsilon_{1} \Delta \epsilon_{2}$. In particular, for the states with $\Delta \epsilon_{1} \Delta \epsilon_{2}=-1$ the decoherence and the orthogonalization processes are more efficient, again bringing this part of $\rho_{S:fE}(t)$  closer to SBS, than for those with $\Delta \epsilon_{1} \Delta \epsilon_{2}=+1$. However, the asymptotic values for both cases are still higher than that for the single qubit case. Increasing the value of the transit time to $\Lambda \tau = 5$, the differences in the time behavior of both parities $\Delta \epsilon_{1} \Delta \epsilon_{2}=\pm 1$ almost disappear apart from the region around the transit time.  

Next, we investigate the influence of the Ohmicity parameter $s$ (cf. Eq. (\ref{sd})) . In Fig. \ref{fig:dbs} we present the behavior of decoherence and fidelity factors for $s=2$ and $s=3$. This corresponds to a well known transition between the Markovian and non-Markovian evolution of the single spin model and manifests in the change from monotonic to non-monotonic behavior of the single spin decoherence curve \cite{Addis2014,BreuerRev2016}. In the case of the spin register, one can see that already for $s=2$ the decoherence curves are non-monotonic (for off-diagonal elements with $\Delta \epsilon_{1} \Delta \epsilon_{2}=\pm1$) due to the relaxation process around $t=\tau$, and there is no qualitative change between $s=2$ and $s=3$. Comparing Figs. \ref{fig:dbs} and \ref{fig:dbuncut} one sees that the peak/dip becomes more pronounced with increasing the Ohmicity parameter, but the differences between the decoherence and the fidelity curves disappear for the chosen temperature $T=\Lambda/3$. Finally, let us mention that a relation between non-Markovianity and efficiency of SBS formation was studied for a single spin model in \cite{my} and no direct connection between the two processes was found.  Let us also mention that a quantification of non-Markovianity here would require introduction of an appropriate, for the studied model, non-Markovianity measure \cite{Hall2014}.

We now move to the cut spectral density case (cf. \cite{my}). We assume that the observed frequencies belong to a window $[\alpha / \Lambda, (\alpha + \Delta)/\Lambda]$ and the complement of this interval is not observed (the traced out part of the bosonic environment).  In Fig. \ref{fig:dbcut} we present time-asymptotic ($\Lambda t \gg 1$) values of decoherence and fidelity factors as functions of the cut position $\alpha$ and for different  transit times. 
Although the behavior is much more complicated than for a single spin, one can still see a characteristic reciprocal behavior \cite{my} between decoherence (upper plots) and fidelity (lower plots) factors, reflecting reciprocal dependence on the temperature of the two functions.
In addition, one observes small oscillations of both fidelity and decoherence factor for states with $\Delta \epsilon_{1} \Delta \epsilon_{2}= \pm 1$ with respect to the placement of the cut. We verified that for higher values of transit time these oscillations vanish, so that there is no difference in decoherence and fidelity between states with $\Delta \epsilon_{1} \Delta \epsilon_{2}=\pm 1$. For transit times $\Lambda \tau \gg 1$ the "single qubit" case results in weaker decoherence and information transfer to the environment than the other two discussed cases.

\section{Collective decoherence and orthogonalization}\label{s5}
\label{sec:colletive}
Looking at the exact expressions for the uncut (the whole environment traced out) case in Appendix \ref{sec:appendixa}, one sees that the transit times $\tau_{nm}$ always scale with other time-constants: The cut-off frequency $\Lambda$ and the thermal time $\tau_T = 1/(k_BT)$.
Let us now consider the situation when $\tau_{mn}$ are the shortest time-scales in the model:
\be
\tau_{nm}\ll\tau_T,\Lambda^{-1}
\ee
for all $n,m$. In particular, the last condition is equivalent through (\ref{transit}) to that of the qubit separation being much smaller than the wavelengths involved, $\boldsymbol{k}\Delta \boldsymbol{r}_{nm} \ll 1$, or that the coupling constants do not depend on the qubit positions:
\be
g_{\boldsymbol k}^n\equiv g_{\boldsymbol k}
\ee
%i.e. when the phase of the field does not change on the length scale of the register, we find
From (\ref{eq:decmat}) and (\ref{fidelity}) it follows that:

\ben
&&-\log \Gamma_{ \boldsymbol{\epsilon} \boldsymbol{\epsilon}'}(t) = \boldsymbol \Gamma_{11}(t) \left( \sum_n\Delta \epsilon_n \right)^2 \\
&&+ i \boldsymbol \Gamma^+_{11}(t) \left[ \left( \sum_n \epsilon_n \right)^2 - \left( \sum_n \epsilon_n' \right)^2 \right]
\een
and
\ben
-\log B^{{\cal M}}_{\boldsymbol{\epsilon} \boldsymbol{\epsilon}'}(t) =\boldsymbol B^{{\cal M}}_{11}(t) \left( \sum_n\Delta \epsilon_n \right)^2
\een

The quantities $ \boldsymbol \Gamma_{11}(t), \boldsymbol B^{{\cal M}}_{11}(t)$ are just the single-qubit ($L$=1) decoherence and distinguishability factors, analyzed e.g. in \cite{my}. Hence, in this regime the whole register behaves almost like  a collection of independent qubits, with the important qualitative difference of existence of DFS and OFS (see Section \ref{subsec:subspaces}) . In particular, a simultaneous strong DFS and OFS occur  for all pairs $\boldsymbol{\epsilon}, \boldsymbol{\epsilon}'$ such that (cf. \cite{RQJ2002}):
\ben
\label{eq:DFS}
&&\sum_n\Delta \boldsymbol{\epsilon}_n  = \sum_n (\boldsymbol{\epsilon}_n-\boldsymbol{\epsilon}_n')=0,  \label{wDFS} \\
&& \left( \sum_n \boldsymbol{\epsilon}_n \right)^2 - \left( \sum_n \boldsymbol{\epsilon}_n' \right)^2=0 \label{sDFS}
\een
while a simultaneous weak DFS and OFS occurs when only (\ref{wDFS}) is fulfilled. It is interesting that the same condition (\ref{wDFS}) controls both decoherence and state fidelity.
 An example of a simultaneous strong DFS and OFS is  the
subspace of a $2$-qubit register spanned by $|\frac{1}{2},-\frac{1}{2}\rangle$ and $|-\frac{1}{2},\frac{1}{2}\rangle$  (analyzed in more detail in \cite{RQJ2002}).
On the other hand, the pair of states that decohere most and become most distinguishable are those for which  $\sum_n\Delta \boldsymbol{\epsilon}_n$ is the largest.

Let us investigate the structure of the partially reduced state in the presence of a simultaneous strong DFS and OFS. Let us denote this subspace as $DFS$ and assume it to be strong. For the sake of clarity, we consider it to be two-dimensional, spanned by vectors $\tilde{\boldsymbol{\epsilon}}, \tilde{\boldsymbol{\epsilon}}'$ (the extension to higher dimensions is analogous). We find that (cf. Eq. (\ref{eq:controlunitary}))
 \ben \hat U_{\boldsymbol{k}} (t;\tilde{\boldsymbol{\epsilon}} ) = \hat U_{\boldsymbol{k}} (t;\tilde{\boldsymbol{\epsilon}}' ),
 \een
which be immediately verified using Eqs. (\ref{eq:DFS}, \ref{sDFS}):
\ben
&&U^I_{\boldsymbol{k}}(t;\boldsymbol{\tilde{\epsilon}} )^{\dagger}U^I_{\boldsymbol{k}}(t;\boldsymbol{\tilde{\epsilon}}) = \\  &&\hat D\left( \alpha_{\boldsymbol{k}}(t)  g_{\boldsymbol{k}} \sum_n \Delta \boldsymbol{\epsilon}_n   \right)  e^{i g_{\boldsymbol{k}}\left[ \left( \sum_n \boldsymbol{\epsilon}_n \right)^2 - \left( \sum_n \boldsymbol{\epsilon}_n' \right)^2 \right] \xi_{\boldsymbol{k}} (t) } = \nonumber \\&& I. \nonumber
\een
As a result the controlled unitary operator has a form:
\ben
&&\hat U^I_{S:E}(t)=  \hat \Pi_{DFS}  \ot  \bigotimes_{\boldsymbol{k}}^{fN} \hat U^I_{\boldsymbol{k}}(t;DFS ) + \nonumber \\ &&\sum_{\boldsymbol{\epsilon} \notin DFS}| \boldsymbol{\epsilon} \> \< \boldsymbol{\epsilon} |  \ot  \bigotimes_{\boldsymbol{k}}^{fN} \hat U^I_{\boldsymbol{k}}(t;\boldsymbol{\epsilon} ) ,
\een
where $\hat \Pi_{DFS}$ is a projector onto $DFS$, i.e.
\ben
\hat \Pi_{DFS} = |\tilde{\boldsymbol{\epsilon}} \> \< \tilde{\boldsymbol{\epsilon}} | + | \tilde{\boldsymbol{\epsilon}}' \> \< \tilde{\boldsymbol{\epsilon}}' | +| \tilde{\boldsymbol{\epsilon}} \> \< \tilde{\boldsymbol{\epsilon}}' | + | \tilde{\boldsymbol{\epsilon}}' \> \< \tilde{\boldsymbol{\epsilon}} |.
\een
Let us assume that the conditions for formation of SBS are fulfilled for register states not belonging to $DFS$, then the partially reduced state  is
\ben
\label{eq:colectiveSBS}
&&\rho^I_{S:fE}=\hat \Pi_{DFS} \rho_{0S} \hat \Pi_{DFS} \otimes \bigotimes_{\cal M}^{f \cal M} \rho^I_{\cal M}(t; DFS) \nonumber + \\ &&
\sum_{\boldsymbol{\epsilon} \notin DFS } c_{\boldsymbol{\epsilon},\boldsymbol{\epsilon}}| \boldsymbol{\epsilon} \> \< \boldsymbol{\epsilon} | \otimes \bigotimes_{\cal M}^{f \cal M} \rho^I_{\cal M}(t; \boldsymbol{\epsilon}).
\een
This is what we call a coarse-grained  SBS \cite{Mironowicz2017}: The SBS structure is build upon subspaces rather than states (the pointer states) and coherences are in general preserved within the subspaces and the information leaked into the environment allows to discriminate only between subspaces but not between the vectors they are spanned on. Further studies on various forms of departure from SBS can be found in \cite{Mironowicz2017}.

\section{Conclusions}\label{s6}

We revisited decoherence process of a multi-qubit register interacting with a bosonic thermal bath. Unlike in the previous studies \cite{PSE1996,RQJ2002} we were interested in information gained by the environment. To this end, we employed a recently introduced notion of Spectrum Broadcast Structures (SBS) \cite{Korbicz2014, myPRA}, which are specific  multipartite quantum state structures describing redundant encoding of system information (here the register state) in the environment.  We explicitly calculated the relevant figures of merit describing the SBS formation -- the usual decoherence factor and mixed state fidelities -- in the simple case of so called uncut environment, where each observer observing the environment has an access to the full frequency spectrum. Studying more in detail the simplest case of a $2$-qubit register, we pointed out to the causal disturbance propagation between the qubits, which can de- or re-cohere the register state, depending on its parity. Although this was previously known at the level of the register state \cite{Milonni,RzazewskiZakowicz,RQJ2002}, here we showed that there is an accompanying impulse in the environment causing increase/decrease of environment information respectively. In another simple case of a collective decoherence, corresponding to vanishingly small transit times of the field disturbance between the qubits, we showed a coarse grained SBS. These are quite interesting structures, appearing to the presence of protected spaces.  

The model considered here was quite simple with a trivial dynamics of the register. One future direction would be studies of a more realistic full model, which includes register tunneling \cite{Milonni,RzazewskiZakowicz}. However, already at the sole central system level the dynamics is rich and complicated, e.g. with multiple causal impulses propagating between the qubits. 

Moreover, let us further elaborate on the link between dynamics of multi-qubit register and Dicke supperradiance. The standard picture of decay of unstable states, such as occurs in Dicke superradiance, is that the first radiated photons are completely spontaneous and random, then the signal amplifies and becomes classical -- e.g. it can be described to a high degree of accuracy by coherent states \cite{Arecchi_AtomicCoherentStates}, closely resembling properties of classical states. This way of achieving "classicality" might probably be a mechanism of SBS formation. Classical states achieved in such processes are very random since they result from amplification of the spontaneous initial part of radiation. Therefore, this process is usually regarded as a manifestation of macroscopic quantum fluctuations \cite{Haake,Haake1979}. It would be very interesting to use multi-qubit register to investigate this problem form SBS perspective, and we leave this for a further study.

\begin{acknowledgments}
We would like to thank \L. Cywi\'nski, M. Ku\'s and P. Horodecki for discussions. The work was made possible through the support of
grant ID \# 56033 from the John Templeton Foundation. M.L. and A.L acknowledge the Spanish Ministry MINECO (National Plan
15 Grant: FISICATEAMO No. FIS2016-79508-P, SEVERO OCHOA No. SEV-2015-0522, FPI), European Social Fund, Fundació Cellex, Generalitat de Catalunya (AGAUR Grant No. 2017 SGR 1341 and CERCA/Program), ERC AdG OSYRIS, EU FETPRO QUIC, and the National Science Centre, Poland-Symfonia Grant No. 2016/20/W/ST4/00314. J.T  is supported by
the Swedish Research Council under Contract No. 335-2014-7424.

\end{acknowledgments}
\appendix
\section{Uncut environment - analytical formulas}
\label{sec:appendixa}
In this appendix we provide analytical expression for decoherence factor and mixed state fidelity. As in the main text we assume that the spectral density is given by the following expression
\ben
J(\omega) = \frac{\omega^s}{\Lambda^{s-1}}\exp \left[-\omega/\Lambda\right],
\een
with $s>1$.
As usually \cite{SchlosshauerBook,BreuerBook} decoherence factor factorizes into the vacuum and thermal parts, $\boldsymbol{\Gamma}(t)=\boldsymbol{\Gamma}^{vac}(t)+\boldsymbol{\Gamma}^{th}(t)$, which in the considered model are:

\ben
&&\boldsymbol{\Gamma}_{nm}^{vac}(t)= \label{Gvac} \\ \nonumber &&\frac{\wp(s-1)}{2} \left[2  (1+(\Lambda \tau_{nm})^2)^{\frac{1-s}{2}}  \right.   \cos \left[(s-1) \arctan \left(\Lambda \tau_{nm}\right)\right] - \\ \nonumber && - (1+(\Lambda \boldsymbol{t}^-_{nm})^2)^{\frac{1-s}{2}}    \cos \left[(s-1) \arctan \left(\Lambda \boldsymbol{t}^-_{nm}\right)\right] - \\ \nonumber && -(1+(\Lambda \boldsymbol{t}^+_{nm})^2)^{\frac{1-s}{2}} \left.    \cos \left[(s-1) \arctan \left(\Lambda\boldsymbol{t}^+_{nm}\right)\right] \vphantom{\frac{1}{2^5}}
 \right] \nonumber \\
&&\boldsymbol{\Gamma}^{th}_{nm}(t)=\label{Gthm} \\ & \nonumber &\frac{(-1)^{s-1}}{(\Lambda \tau_T)^{s-1}} \left[ 2\Psi^{(s-2)} \left(1+ \frac{1}{\Lambda \tau_T} + i\frac{ \tau_{nm}}{ \tau_T}\right) \right. \\ \nonumber  && - \Psi^{(s-2)} \left( 1+ \frac{1}{\Lambda \tau_T} -  \frac{i\boldsymbol{t}^+_{nm}}{\tau_T}  \right) \\ \nonumber &&\left. - \Psi^{(s-2)} \left( 1+ \frac{1}{\Lambda \tau_T} -  \frac{i\boldsymbol{t}^-_{nm}}{\tau_T}  \right) + c.c. \right]
\een
where $\Psi^m(z)$ is the so-called polygamma function defined as \cite{NIST}:
\ben
\Psi^{m}(z) \equiv \frac{d^{m+1}}{dz^{m+1}} \ln \wp (z) =  \sum_{k=0}^{\infty} \frac{(-1)^{m+1} m!}{(z+k)^{m+1}},
\een
$\wp(z)$ is the Euler gamma function, $C.c$ denotes complex conjugated part  and we introduced advanced and retarded times:
\ben
\label{eq:transit}
&&\boldsymbol{t}^\pm_{nm} \equiv t \pm \tau_{nm},
\een
and $\tau_T = 1/(k_BT)$.
The quantities entering the phases are
\ben
\label{eq:decan}
&& \boldsymbol{\Gamma}^+_{nm}(t) =  \frac{\wp(s-1)}{2} \times\\ && \left[2 (s-1) \Lambda t (1+ (\Lambda \tau_{nm})^2)^{-\frac{s}{2}}\cos \left[s \arctan (\Lambda \tau_{nm})\right] \right. \nonumber \\ && - (1+(\Lambda \boldsymbol{t}^-_{nm})^2)^{\frac{1-s}{2}}  \sin \left[(s-1) \arctan (\Lambda \boldsymbol{t}^-_{nm})\right] \nonumber   \\ &&\left. -(1+(\Lambda\boldsymbol{t}^+_{nm})^2)^{\frac{1-s}{2}} \sin \left[(s-1) \arctan (\Lambda\boldsymbol{t}^+_{nm})\right]\right] \nonumber \\
&&\boldsymbol{\Gamma}^-_{nm}(t) = \\ \nonumber &&  \frac{\wp(s-1)}{2} \left[2 (1+(\Lambda \tau_{nm})^2)^{\frac{1-s}{2}}\sin \left[(s-1) \arctan (\Lambda t_{nm})\right] \right. \\ \nonumber &&+ (1+(\Lambda \boldsymbol{t}^-_{nm})^2)^{\frac{1-s}{2}} \sin \left[(s-1) \arctan (\Lambda \boldsymbol{t}^-_{nm} )\right] \nonumber  \\ \nonumber &&\left. + (1+(\Lambda \boldsymbol{t}^+_{nm})^2)^{\frac{1-s}{2}} \sin \left[(s-1) \arctan (\Lambda\boldsymbol{t}^+_{nm})\right]\right], \nonumber
\een

Distinguishability can also be decomposed into vacuum and thermal part $\boldsymbol B^{{\cal M}}(t) = \boldsymbol B^{{\cal M} \; vac}(t)+\boldsymbol B^{{\cal M} \; th}(t)$, with $\boldsymbol B^{{\cal M} \; vac}(t) = \boldsymbol \Gamma^{vac}(t)$ and
\ben
\label{eq:fidan}
&&\boldsymbol B^{{\cal M} \; th}_{nm}(t) =\\ & \nonumber &\frac{(-1)^{s-1}}{(2\Lambda \tau_T)^{s-1}} \left[ \Psi^{(s-2)} \left(1+ \frac{1}{2\Lambda \tau_T} + \frac{i \tau_{nm}}{2 \tau_T}\right) \right. \\ \nonumber &&-\Psi^{(s-2)} \left(\frac{1}{2}+ \frac{1}{2\Lambda \tau_T} + \frac{i t_{nm}}{2 \tau_T}\right) \\ \nonumber -  &&\frac{1}{2}\Psi^{(s-2)} \left(1+ \frac{1}{2\Lambda \tau_T} - \frac{i \boldsymbol{t}^+_{nm}}{2 \tau_T}\right)\\ \nonumber \nonumber &&+\frac{1}{2}\Psi^{(s-2)} \left(\frac{1}{2}+ \frac{1}{2\Lambda \tau_T} - \frac{i \boldsymbol{t}^+_{nm}}{2 \tau_T}\right)\\ \nonumber-  &&\frac{1}{2}\Psi^{(s-2)} \left(1+ \frac{1}{2\Lambda \tau_T} - \frac{i \boldsymbol{t}^-_{nm}}{2 \tau_T}\right)\\ \nonumber \nonumber && \left. +\frac{1}{2}\Psi^{(s-2)} \left(\frac{1}{2}+ \frac{1}{2\Lambda \tau_T} - \frac{i \boldsymbol{t}^-_{nm}}{2 \tau_T}\right)  + C.c. \right]
\een

\section{Superradiance and radiation trapping}
\label{sec:appendixb}

In 1954 Dicke \cite{Dicke54} predicted that $N$ emitters/dipoles will radiate collectively
with the rate proportional to $N$ and intensity proportional to $N^2$, provided they are confined in a region of the size much smaller than $\lambda^D$ in D dimensions, where $\lambda$ is the wave length of the radiation. Only one mode of such systems, the one in which all dipoles are parallel and have the same phase, will exhibit collective superradiance. All the other modes, with the total dipole moment equal to zero, will be non-radiative, or in practice will radiate  very slowly. Since the famous paper of Dicke, a lot of work has been done on collective
emission from a system of many sources. In the 1970s and 1980s the theory focused more on superflourescence, and considered
usually  the pencil-shape samples, of dimensions large compared with the
wavelength, for which propagation effects play a dominant role (e.g.\cite{Banfi,Haa71,Haake}). It  was  widely believed that in the small-sample limit the original Dicke's
description is valid. Namely, in such a limit, only a global dipole moment is coupled to
the radiation; the lifetime of the excitation of this degree of freedom is $N$ times shorter
than the lifetime of a single atom and the excitations of all the other global modes of the
system are trapped and cannot decay through radiative damping. This simple picture
can be inadequate due to near zone effects, as was suggested in the paper of Friedberg and Hartman \cite{Friedberg1974a,Friedberg1974b}
which dealt with a small spherical sample. This problem was studied for the
spherical sample composed of charged harmonic oscillators by Zakowicz (1978).
Essential for these studies was taking into account the near-zone field in the system, and
longitudinal dipole-dipole forces in particular.

On the other hand, there are a lot of papers studying the problem of two atoms \cite{Agarwal,Kus,Milonni}.
In fact, it is known from the  paper of Stephen \cite{Stephen}  that proper collective broadening and
narrowing of the emission line conform to a simple picture developed by Dicke. In  contrast, in Ref. \cite{Lewenstein}, it was shown, that Dicke's  superradiance, strictly speaking,  cease to exist
in a system of 4 atoms (harmonic oscillators) located on the vertices of a tetrahedron or equilateral triangle.  This effect results from  the mode mixing due to the strong near zone interactions at short distances, smaller than the wave length.

Recently, there has been a revival of interest in radiation trapping and collective emission in the context of quantum nanophotonics \cite{Chang}. Here, the mechanisms of radiation trapping are more subtle and are governed by destructive interference patters.

\bibliography{Objectivity}

\end{document}